\documentclass{article}
\usepackage{spconf,graphicx}
\usepackage{amsmath,amssymb}
\usepackage{mathtools}
\usepackage{csquotes}
\usepackage{lipsum}
\usepackage{siunitx}
\usepackage{booktabs,tabularx}
\usepackage{pifont}
\usepackage[backend=biber,
            style=ieee,
            doi=false,
            isbn=false,
            url=false]{biblatex}
\usepackage[keeplastbox]{flushend}
\usepackage{caption}
\usepackage{microtype}
\usepackage{tikz}
\usetikzlibrary{positioning,backgrounds,fit,dsp}
\usepackage[hidelinks]{hyperref}
\usepackage[capitalise]{cleveref}
\newcommand{\pA}[0]{P_\mathcal{A}}
\newcommand{\pC}[0]{P_\mathcal{C}}
\newcommand{\fA}[0]{f_{\mathcal{A}}}
\newcommand{\fC}[0]{f_{\mathcal{C}}}
\newcommand{\A}[0]{\mathbf{A}}
\newcommand{\X}[0]{\mathbf{X}}
\newcommand{\Xstar}[0]{\mathbf{X}^{\text{\ding{72}}}}
\newcommand{\Y}[0]{\mathbf{Y}}
\newcommand{\Z}[0]{\mathbf{Z}}
\newcommand{\DZest}[0]{\Delta\Z^{\text{est}}}
\newcommand{\DZestm}[0]{\Delta\Z^{\text{est},m}}
\newcommand{\DIPcg}[0]{\text{DIP}_{cg}}
\newcommand{\DIPvm}[0]{\text{DIP}_{vm}}
\newcommand{\DIPDM}[0]{\text{DIP}^{\text{DM}}}
\newcommand{\norm}[1]{\left\lVert#1\right\rVert}
\newcommand{\minitimes}{{\mkern-2mu\times\mkern-2mu}}
\DeclareMathOperator{\STFT}{STFT}
\DeclareMathOperator{\iSTFT}{iSTFT}

\DeclareSIUnit{\px}{\text{px}}
\title{Deep Iterative Phase Retrieval for Ptychography}
\name{Simon Welker$^{1,2}$, Tal Peer$^{1}$, Henry N. Chapman$^{2}$, Timo Gerkmann$^{1}$\thanks{\scriptsize We acknowledge the support by DASHH (Data Science in Hamburg - HELMHOLTZ Graduate School for the Structure of Matter) with the Grant-No. HIDSS-0002 and by the Deutsche Forschungsgemeinschaft (DFG, German Research Foundation) project number 247465126.}}
\address{$^{1}$ Signal Processing (SP), Universität Hamburg, Germany \\
      $^{2}$ Center for Free-Electron Laser Science, DESY, Hamburg, Germany}  %
\begin{document}
\ninept
\maketitle
\begin{abstract}
One of the most prominent challenges in the field of diffractive imaging is the phase retrieval (PR) problem: In order to reconstruct an object from its diffraction pattern, the inverse Fourier transform must be computed. This is only possible given the full complex-valued diffraction data, i.e. magnitude and phase. However, in diffractive imaging, generally only magnitudes can be directly measured while the phase needs to be estimated. In this work we specifically consider ptychography, a sub-field of diffractive imaging, where objects are reconstructed from multiple overlapping diffraction images. We propose an augmentation of existing iterative phase retrieval algorithms with a neural network designed for refining the result of each iteration. For this purpose we adapt and extend a recently proposed architecture from the speech processing field. Evaluation results show the proposed approach delivers improved convergence rates in terms of both iteration count and algorithm runtime.
\end{abstract}
\section{Introduction}
\vspace{-2mm}
\label{sec:intro} %
Phase retrieval is a problem encountered within several fields which employ techniques based on complex integral transforms (most notably the Fourier transform). While in many cases only the magnitude of a complex signal is available, one also requires its phase in order to perform the transformation from one domain to another. While other approaches exist, estimation of the phase from the magnitude is most commonly performed using iterative methods which alternate between the domains while applying a set of constraints on the signal in each domain \cite{shechtmanPhaseRetrievalApplication2015,fienupReconstructionObjectModulus1978,griffinSignalEstimationModified1984,gerkmannPhaseProcessingSingleChannel2015}.

In this work we consider the phase retrieval problem in the otherwise unrelated fields of diffractive X-ray imaging and speech signal processing, and seek to adapt recently proposed methods from the latter to the former. In particular, we draw an analogy between the short-time Fourier transform (STFT) and ptychography and transfer the recently proposed Deep Griffin-Lim Iteration \cite{masuyamaDeepGriffinLim2019,masuyamaDeepGriffinLim2021} into the ptychography domain. This approach augments a conventional iterative approach (Griffin-Lim algorithm \cite{griffinSignalEstimationModified1984}) with a convolutional neural network, leading not only to an improved phase reconstruction accuracy, but also to better convergence rates.

Exploiting the conceptual similarity between ptychography and STFT, as well as the similarity between the iterative phase retrieval algorithms used in both domains, we propose Deep Iterative Projections (DIP) --- an augmented iterative approach to phase retrieval for ptychography applications.

\subsection{Related Work}
\vspace{-7px}
PtychNet \cite{kappelerPtychnetCNNBased2017} is a CNN for the direct reconstruction of Fourier ptychography data, and SspNet \cite{wengrowiczDeepNeuralNetworks2020} is an encoder-decoder DNN architecture for single-shot ptychography (SSP). Both PtychNet and SspNet are non-iterative methods and directly reconstruct imaged objects without explicitly reconstructing phase information. Deep Ptych \cite{shamshadDeepPtychSubsampled2019} utilizes generative prior models to regularize the phase problem, showing increased reconstruction quality over classical methods but requiring an applicable prior model. prDeep \cite{metzlerPrDeepRobustPhase2018} is based on the optimization of an explicit minimization objective involving a generic denoising DNN, which is trained for a particular target data distribution. It is therefore closely related to Plug-and-Play approaches; in contrast, the method we present here uses a task-oriented denoising DNN structured around an iterative projection algorithm and does not use an external optimizer. Estupiñań et al. \cite{estupinanDeepUnrolledPhase2021} propose PR from coded diffraction patterns with an unrolled DNN approach, thereby fixing the number of iterations for retrieval in the DNN architecture. Jagatap and Hegde \cite{jagatap2019phase} apply an untrained network priors approach to a generic non-ptychographic PR problem. An approach closest to ours \cite{isilDeepIterativeReconstruction2019} combines two separate DNNs, one of them also trained to improve upon a classical phase retrieval algorithm (Hybrid input-output \cite{fienupPhaseRetrievalAlgorithms1982}), but is designed for phase retrieval from single diffraction patterns rather than ptychography data.
\begin{figure}[b]
    \centering
    \scalebox{0.9}{  \tikzstyle{normalnode}=[dspsquare, inner sep=1mm]%
    \tikzstyle{smallnode}=[normalnode, minimum height=0.5cm, minimum width=0.8cm]
    \tikzstyle{connarrow}=[dspconn]  

    \begin{tikzpicture}[node distance=1cm and 1.3cm]
        \node (Xin) {};
        \node[dspadder, right=0.8cm of Xin] (noise_adder) {};
        \node [above=0.6cm of noise_adder] (noise_in) {};
        \node[smallnode, right=of noise_adder] (PA) {$\pA$};
        \node[smallnode, right=of PA] (PC) {$\pC$};
        \node[normalnode, right=1cm of PC, yshift=-0.8cm, fill=green!30, minimum height=1cm] (DNN) {DNN};
        \node[dspadder, right=of DNN, xshift=-1.2cm, yshift=0.8cm] (adder) {};
        \node[right=1cm of adder] (Xout) {};

        \node[above=0.5cm of PA] (Ain_PA) {};
        \node[left=0.5cm of DNN,yshift=-0.3cm] (Ain_DNN) {};

        \draw[connarrow] (Xin) -- (noise_adder);
        \draw[connarrow] (noise_in) -- node [right] {$\varepsilon$} (noise_adder);
        \draw[connarrow] (noise_adder) -- node[dspnodefull] (Xin_split) {$\X^m$} (PA);
        \draw[connarrow] (PA) -- node[dspnodefull] (PA_split) {$\Y^m$} (PC);
        \draw[connarrow] (PC) -- node[dspnodefull, xshift=-0.4cm] (PC_split) {$\Z^m$} (adder);

        \draw[connarrow] (Xin_split) |- ([yshift=-0.1cm]DNN.west);
        \draw[connarrow] (PA_split) |- ([yshift=0.1cm]DNN.west);
        \draw[connarrow] (PC_split) |- ([yshift=0.3cm]DNN.west);
        \draw[connarrow] (DNN.east) -| node[right,yshift=0.5cm] {$-$} (adder);
        \draw[connarrow] (adder) -- node [dspnodefull, minimum size=0] {$\X^{m+1}$} (Xout); %
        
        \draw[connarrow, dotted] (Ain_PA) -- node [right] {$\A$}  (PA);
        \draw[connarrow, dotted] (Ain_DNN) -- node [below] {$\A$} ([yshift=-0.3cm]DNN.west);
        
        \begin{pgfonlayer}{background}

        \node[behind path, draw,
        rectangle, 
        fill=orange!30,
        rounded corners=1mm, 
        dashed, 
        fit=(PA) (PC) (Ain_PA)
        ]
    (AP_rect) {};
        \node[anchor=north east, inner sep=3pt] at(AP_rect.north east) {\footnotesize AP};
        
        \node[behind path, draw,
            rectangle, 
            fill=blue!30,
            rounded corners=1mm, 
            dashed, 
            minimum width=1cm, inner sep=0, 
            fit=(noise_in) (noise_adder) (AP_rect.north -| noise_adder.north) (AP_rect.south -| noise_adder.south),
            label=below:{\footnotesize (At training)}
            ]
        (noise_train) {};
        \end{pgfonlayer}

    \end{tikzpicture}}
    \caption{Architecture diagram of a single iteration of phase retrieval with DIP. The estimated ptychograph $\X^m$ is subjected to the two projections $\pA$ and $\pC$ making up the alternating projections (AP) algorithm. It is fed into a DNN along with the intermediate results $\Y^m, \Z^m$ and the known Fourier amplitudes $\A$, which estimates a residual ptychograph. During training, random noise $\varepsilon$ is added to clean ptychographs and the loss between the estimated and true residual is minimized.}
    \label{fig:architecture}
\end{figure}
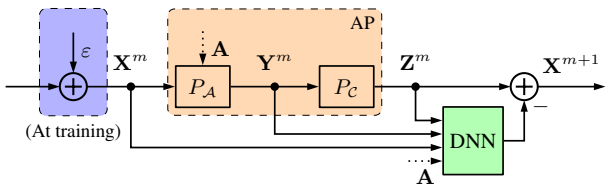
\section{Iterative Phase Retrieval}
\vspace{-2mm}
\label{sec:phase_retrieval}
Iterative phase retrieval techniques aim to recover the phase information required for a successful reconstruction from data like measured diffraction patterns or the STFT of an audio signal, by applying a set of constraints in an alternating and iterative fashion. It is a fairly general technique for solving the phase problem, with a variety of possible constraints to choose from \cite{maidenImprovedPtychographicalPhase2009, elserPhaseRetrievalIterated2003}. In this paper, we will only consider the two most typical projections for ptychographic phase retrieval, namely the \emph{amplitude constraint} and the \emph{consistency constraint}. These are---not coincidentally---also the same constraints applied for iterative phase retrieval in speech processing \cite{griffinSignalEstimationModified1984,gerkmannPhaseProcessingSingleChannel2015}. The amplitude constraint enforces the measured amplitudes in the Fourier domain, modifying the amplitudes in a spectrogram while leaving the phases unchanged. We can enforce it via an element-wise projection operation $\pA$:
\begin{equation}
    \pA(\X) = \A\frac{\X}{|\X|+\delta} \hspace{1.5em} \text{with}\ \X \in \mathbb{C}^{\mathbf{D}}, \A \in \mathbb{R}^{\mathbf{D}}
\end{equation}
where $\X$ is a complex spectrogram, $\A$ is the known amplitude spectrogram, $\mathbf{D} = (K \minitimes M)$ for spectrograms and $\mathbf{D} = (K \minitimes L \minitimes M \minitimes N)$ for ptychographs, and $\delta$ is a small numerical fudge factor to avoid division by zero. The consistency constraint forces overlapping segments to be consistent with each other, by applying an inverse Fourier transform to all segments, summing up these segments to a single object, and applying the known segmentation followed by a forward Fourier transform. Calling this backward transformation \emph{iSTFT} and the forward transformation \emph{STFT} in accordance with speech processing literature, the corresponding projection can be expressed as:
\begin{equation}
    \pC(\X) = \STFT(\iSTFT(\X))
\end{equation}
One should note that the \emph{iSTFT} is, due to the overlap, only a pseudoinverse of the \emph{STFT}, hence $\pC$ is not the identity function. An analogous pair of transformations is utilized in iterative phase retrieval for ptychography, though they are typically considered together as one and called the \emph{object update} \cite{rodenburgPtychography2019}. We can therefore formulate the ptychographic phase retrieval problem as follows, where $\text{Img}(\STFT)$ is the image of the ptychographic forward transformation:
\begin{equation}
    \text{Find }\X\text{ s.t.} \left\{
        \begin{array}{ll}
            |\X| = \A,\\
            \X \in \text{Img}(\STFT)
        \end{array}
      \right.
\end{equation}
The simplest iterative phase retrieval algorithm, called the Griffin-Lim Algorithm (GLA) in speech \cite{griffinSignalEstimationModified1984} and Alternating Projections (AP) in ptychography \cite{rodenburgPtychography2019}, proceeds by repeatedly applying $\pA$ and $\pC$ in an alternating fashion, i.e.,
\begin{equation}
    \X^{m+1} := \pC(\pA(\X^m))
\end{equation}
where $\X^m$ is the spectrogram after the $m$-th iteration. The Difference Map algorithm \cite{elserPhaseRetrievalIterated2003,elserSearchingIteratedMaps2007} is another iterative algorithm for phase retrieval, noted to have realized \enquote{a significant improvement in image quality and resolution over earlier work} \cite{rodenburgPtychography2019}. It is based on the same two projections:
\begin{align}
    f_A(\X) :=&\ \pA(\X) - (\pA(\X) - \X) / \beta \\
    f_C(\X) :=&\ \pC(\X) + (\pC(\X) - \X) / \beta \\
    \X^{m+1} :=&\ \X^m + \beta\left(\pA(\fC(\X^m)) - \pC(\fA(\X^m))\right)
\end{align}
The free parameter $\beta \in [0, 1]$ is typically set to values close to 1. There are other related techniques like PIE and ePIE, which iteratively make changes to single segments rather than the full ptychographs at once, but we will not consider them here. An excellent and detailed overview of this whole class of algorithms can be found in \cite{rodenburgPtychography2019}.
\section{STFT vs. Ptychography}
\vspace{-2mm}
\label{sec:pty}
The \emph{short-time Fourier transform} (STFT) is a widely used technique in speech and audio processing \cite{gerkmannPhaseProcessingSingleChannel2015,benestySpeechEnhancementSTFT2012}. Although a digital audio recording is a one-dimensional time-domain signal, analysis and processing are often carried out in the frequency domain where the signals have favorable properties (e.g. sparsity). However, since the signal is in general non-stationary, simply applying the Fourier transform on the entire signal is not sufficient. A more useful representation is achieved by transforming short overlapping segments, resulting in a two-dimensional complex-valued \emph{time-frequency} representation, commonly referred to as a (complex) spectrogram. A real-valued tapered window function \cite{prabhuWindowFunctionsTheir2018} is applied to the segments before transformation in order to mitigate the leakage effect at the cost of reduced spectral resolution \cite{varyDigitalSpeechTransmission2006}.
Since the measured signal is in the real-valued time-domain, its complex-valued time-frequency representation is perfectly known, i.e. both its magnitude and its phase are available. However, there is still need for phase retrieval in this context. For instance, many speech enhancement algorithms estimate only the clean speech magnitudes but not the noisy phase \cite{wangUnimportancePhaseSpeech1982,paliwalImportancePhaseSpeech2011,varyNoiseSuppressionSpectral1985}. For reconstruction of the time-domain signal, one may either use the unaltered noisy phase, or apply a phase retrieval algorithm to estimate a \emph{consistent} phase \cite{gerkmannPhaseProcessingSingleChannel2015} from the modified magnitude (it is also possible to enhance the noisy phase, which is outside the scope of this work). Another scenario where phase retrieval is required is speech synthesis, where typically only the magnitudes are directly synthesized and the phase is missing \cite{sharmaFastGriffinLim2020}.

In imaging, ptychography refers to a family of computational imaging techniques, where the informational redundancy of recording an object at multiple overlapping positions is effectively exploited to avoid difficult ambiguities that occur when using only a single diffraction pattern. For visible light, and radiation of even shorter wavelength such as X-rays, it is impractical or even impossible to directly record phase information. Recorded diffraction patterns therefore only contain amplitude information of the Fourier components.

The so-called \emph{probe function} describes the shape and phase structure of the beam of radiation used in ptychography. It is the analogue of the STFT's tapered window, albeit with a more complex and often unknown structure. The process of scanning over the sample to record at multiple overlapping positions is analogous to the segmentation process of the STFT, but in ptychography, regular scans are often avoided in practice due to the so-called \emph{grid scan pathology} \cite{thibaultProbeRetrievalPtychographic2009}. The far-field ptychographic process is therefore a close relative of the STFT, but applied in two spatial dimensions rather than one temporal dimension. One could, by analogy, say that the experimental setup in far-field ptychography performs a \emph{local-space 2D Fourier transform}, with the caveat that all phase information is discarded at the time of recording because it cannot be measured directly \cite[Sec. 17.3.1]{rodenburgPtychography2019}.

Here, we will consider the most direct ptychography analogue of the STFT: Far-field ptychography with a known probe and a grid scan pattern that is regular in both spatial directions. Just as the STFT turns a one-dimensional time signal into a two-dimensional time-frequency signal called a \emph{spectrogram}, the corresponding ptychography technique turns a two-dimensional spatial signal --- the exit wave of an imaged 2D object --- into a four-dimensional space$_x$-space$_y$-frequency$_x$-frequency$_y$ signal, which will be referred to in the following as a \emph{ptychograph}.

\section{Proposed Architecture}
\vspace{-2mm}
\label{sec:proposed}
Our proposed neural network architecture for phase retrieval in ptychography, DIP, is based on the DeGLI architecture for speech phase retrieval by Masuyama et al. \cite{masuyamaDeepGriffinLim2021, masuyamaDeepGriffinLim2019}. Here, we first describe DeGLI and then list the adjustments we make to apply it to ptychography data. DeGLI is a deep convolutional neural network (CNN) that aims to augment the Griffin-Lim Algorithm (GLA) \cite{griffinSignalEstimationModified1984} by estimating the residual spectrogram, i.e., the error made by the algorithm. A GLA-like procedure is referred to in ptychography literature as \enquote{alternate projections} \cite{rodenburgPtychography2019}, and is considered the simplest iterative projection algorithm for phase retrieval. The input to the DeGLI network consists of the current estimate of the true complex spectrogram $\X^m$, the known amplitude spectrogram $\A$, as well as $\Y^m=\pA(\X^m)$ and $\Z^m=\pC(\Y^m)$ as defined in \cref{sec:phase_retrieval}. The output of DeGLI is an estimated residual spectrogram $\DZest$, which is subtracted from $\Z^m$ to retrieve a new estimate of the true spectrogram:
\begin{equation}
\label{eq:degli_output}
    \X^{m+1} := \Z^m - \DZestm \, .
\end{equation}

The DeGLI CNN itself consists of several two-dimensional amplitude-gated complex convolutional layers, followed by a $1\minitimes 1$ convolutional layer \cite{masuyamaDeepGriffinLim2021}. The training process of DeGLI follows a denoising strategy: The training samples are complex spectrograms generated from speech recordings, corrupted by additive complex Gaussian noise $\varepsilon$ with a random signal-to-noise ratio  (SNR). The SNR is sampled from a uniform distribution on a chosen interval. The loss function is the squared Frobenius norm between the residual estimated by DeGLI and the residual of the spectrogram $\X$ estimated by GLA compared to the true spectrogram $\Xstar$:
\begin{equation}
\label{eq:degli_loss}
\mathcal{L}_{\text{DeGLI}} = \norm{(\Xstar - \X) - \text{DeGLI}(\X, \Y, \Z, \A)}^2_F \, .
\end{equation}
To apply the DeGLI method to ptychography, we propose several changes: As training data, we generate simulated ptychographs from 2D images rather than spectrograms from 1D speech segments, and replace all 2D complex convolutions with 4D complex convolutions to match the 4D shape of these ptychographs. Since the Frobenius norm in \cref{eq:degli_loss} is a matrix norm, we generalize by summing over all four dimensions.
The authors of DeGLI propose adding the noise component $\varepsilon$ in the training stage to make the network explicitly learn a denoising task. However, here we argue that the network should only learn to denoise the phase (since amplitudes are known). Thus, instead of adding complex Gaussian noise to the complex ptychographs, we propose only adding noise to their phase. For this purpose we use samples from a von-Mises distribution, whose probability density function is given by
\begin{equation}
    \label{eq:vonmises}
    p(\varepsilon) = \frac{\exp\left(\kappa \cos(\varepsilon-\mu)\right)}{2\pi I_0(\kappa)} \, ,
\end{equation}
where $I_0(\cdot)$ is the zeroth-order modified Bessel function of the first kind, $\mu$ is the mean direction and $\kappa$ is a concentration parameter. The von-Mises distribution is the maximum entropy distribution on the circle and hence well-suited for modeling phase noise which is circular by nature \cite{mardiaDirectionalStatistics2000}. Samples are drawn i.i.d. for each ptychograph bin and added to the true phase as shown in \cref{fig:architecture}, with $\mu=0$ and $\kappa$ sampled randomly from a uniform distribution, as for the SNR.
\section{Experimental Setting}
\vspace{-2mm}
\label{sec:exp}
\subsection{Training}
\vspace{-7px}
\label{sec:exp_training}
We set up a simulated phase retrieval experiment on known images. Ptychographs are generated from the $28 \minitimes 28$ training set images in the MNIST \cite{datasetMNIST} dataset, by interpreting the grayscale images as positive real-valued objects and subjecting them to a simulated ptychographic recording process with additive noise as described in \cref{sec:proposed}. We choose a $9 \minitimes 9$ Gaussian probe function with standard deviation $\sigma = 1.5$, and a probe shift of $2$ in both spatial directions. After observing unreasonably large values at the corner pixels during the first iterations of the classical algorithms, we pad the images with one full probe width of zeros to avoid this issue. While we consider all images as purely real-valued objects, we do not enforce this as a constraint in our code, instead using the full 2D FFT everywhere.

For all instances of DIP considered here, we use two inner convolutional layers, as in the original DeGLI design \cite{masuyamaDeepGriffinLim2021}. We set up all convolutions to use $(5 \minitimes 5 \minitimes 3 \minitimes 3)$ kernels, i.e. 5 neighboring diffraction patterns and 3 neighboring frequencies in each spatial direction. Each convolutional layer has 16 channels and no bias terms. To maintain the shape of the input ptychographs, we insert appropriate zero-padding layers after each convolutional layer.
Using the Adam optimizer \cite{kingmaAdamMethodStochastic2017} and training for 300 epochs with a learning rate of $\num{4e-4}$, we train two DIP models: One using the original complex Gaussian noise model proposed for DeGLI \cite{masuyamaDeepGriffinLim2021} ($\DIPcg$) with a random uniformly distributed SNR $\sim \mathcal{U}[\SI{-24}{\dB}, \SI{0}{\dB}]$, and one using our own von-Mises phase-only noise model ($\DIPvm$) with a random $\kappa \sim \mathcal{U}[0.01, 3.0]$. These values were chosen empirically.
\begin{figure}
    \centering
    \includegraphics[width=0.9\columnwidth]{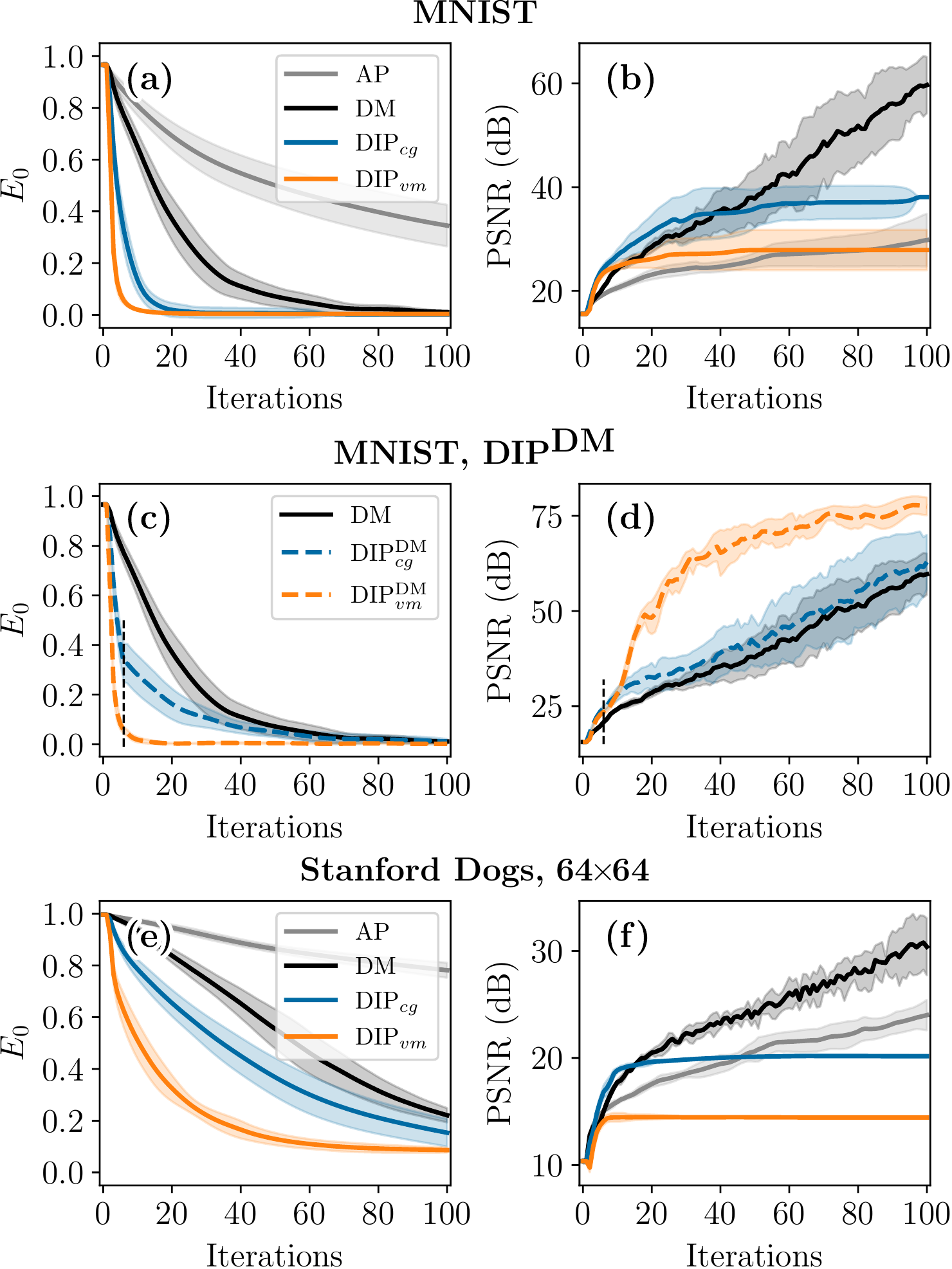}
    \caption{
    Quality achieved by different phase retrieval techniques after 0 to 100 iterations, measured by $E_0$ (left column) and PSNR (right column), averaged over 16 images and 5 random phase initializations each. Filled regions indicate standard deviations.
    \textbf{(a), (b)} On-dataset results, with the same settings and dataset as during training. \textbf{(c), (d)} On-dataset results, evaluating the two DIP instances as initialization steps ($\DIPDM$) that are replaced by DM after 5 iterations (dashed black line). \textbf{(e), (f)} Off-dataset results on Stanford Dogs \cite{datasetStanfordDogs} resized and cropped to $64 \minitimes 64$, with otherwise identical settings.
    }
    \label{fig:errors_overview}
\end{figure}
\subsection{Inference}
\vspace{-7px}
\label{sec:exp_inference}
For evaluation, we let both instances of DIP run iteratively starting from input ptychographs with fully randomized phase information, competing against the simple Alternate Projections (AP) \cite{rodenburgPtychography2019} algorithm---which they aim to augment---as well as the widely used Difference Map (DM) \cite{elserPhaseRetrievalIterated2003,elserSearchingIteratedMaps2007} algorithm with $\beta = 1$. We furthermore evaluate the viability of DIP as an initialization step, by letting DIP run for only 5 initial iterations and then using DM for the remaining iterations ($\DIPDM$). In a slight deviation from DeGLI \cite{masuyamaDeepGriffinLim2019,masuyamaDeepGriffinLim2021}, we do not follow each reconstruction with a final $\pA$ projection, since we found this step to not significantly improve the results of DIP and at the same time to reduce the quality of the reconstructions by DM. We include AP as an initial single step for all competing reconstruction techniques, since we found this to generally improve convergence, in particular for DM. As error measures, we use $E_0$ from ptychography literature \cite{maidenImprovedPtychographicalPhase2009} and the peak signal-to-noise ratio (PSNR) \cite{horeImageQualityMetrics2010} of the reconstructed object amplitudes. $E_0$ is a normalized RMSE metric that includes a correction factor to be invariant to a global scaling factor and phase offset. Both measures require a ground truth object and are therefore only applicable for simulations.

In each scenario, we use a randomly chosen batch of 16 images from an image dataset's test set, from which we generate simulated ptychographs with the same settings as during training. We discard the phase information, re-initializing it randomly, and let each reconstruction technique run for 100 iterations. We repeat this experiment for 5 different random initial phase configurations, evaluate $E_0$ and the PSNR on all reconstructed objects, and calculate the mean and standard deviation over all images in the batch and all initial phases.

\section{Results}
\vspace{-2mm}
\label{sec:results}
\begin{figure}
    \centering
    \includegraphics[width=0.9\columnwidth]{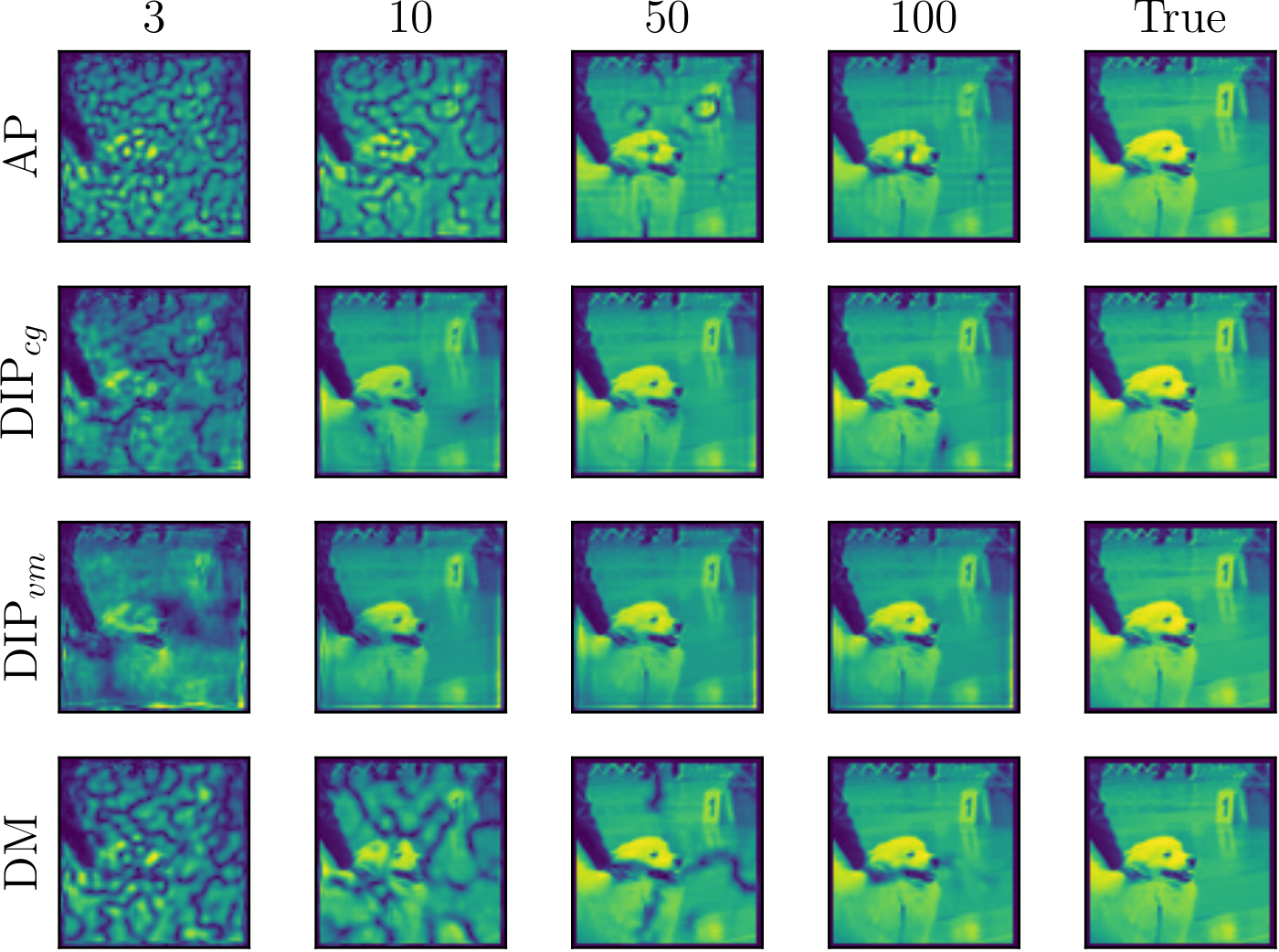}
    \caption{Reconstructions of a single image from the Stanford Dogs dataset. Rows: different reconstruction techniques as in \cref{fig:errors_overview}~\textbf{(e,f)}. Columns: different iteration indices as well as the true image (\emph{True}). Both DIP instances suppress ghostlike artifacts more quickly than the algorithmic techniques. Unlike AP, they also avoid periodic artifacts.}
    \label{fig:images}
\end{figure}
\subsection{On-dataset (MNIST)}
\vspace{-7px}
\label{ssec:on-dataset}
In the first scenario comparing the different reconstruction techniques, we use a random batch of 16 images from the MNIST dataset \cite{datasetMNIST}, simulating ptychographs with the same settings as during training. \cref{fig:errors_overview} (top row) shows how $E_0$ and PSNR of the reconstructed objects evolve with respect to the iteration number.
With respect to $E_0$ (\cref{fig:errors_overview} \textbf{(a)}), both versions of DIP exhibit very fast convergence, showing a marked improvement over AP and even outperforming DM. The quick convergence of our DIP instances is supported by the comparatively high reconstruction quality they achieve during the first few iterations, see \cref{tab:runtime_to_e0}. \cref{tab:runtime_to_e0} also compares the overall runtime required until a $E_0$ value of 0.1 or less is reached. In this scenario, $\DIPvm$ and $\DIPDM_{vm}$ also come out on top, requiring only 5 iterations and less than one second to achieve $E_0 < 0.1$. Furthermore, note that this result is highly implementation-dependent and there is likely room for improvement.

The PSNR values (\cref{fig:errors_overview} \textbf{(b)}) tell, in part, a different story: Both DIP instances stagnate after about 50 iterations, even AP eventually catches up to $\DIPvm$, and DM performs best in this comparison. Therefore, we evaluate the viability of the DIP instances as an initialization step, running each DIP instance for 5 iterations and then switching to DM ($\DIPDM$). The results are shown in the center row of \cref{fig:errors_overview}. The combined technique $\DIPDM_{vm}$ exhibits excellent PSNR values with low variance, and also maintains excellent convergence behavior with respect to $E_0$. We therefore argue that $\DIPDM_{vm}$ performs best in this scenario overall. We posit that this result is due to $\DIPvm$'s ability to particularly quickly eliminate ghost artifacts which seem to otherwise hold up the convergence of DM in this scenario (see \cref{fig:images}), as well as an apparent advantage of DM over DIP in resolving fine detail. We thus also show a clear advantage of our novel proposed training scenario using von-Mises distribution phase-only noise ($\DIPDM_{vm}$) over the original training task using complex Gaussian noise as proposed in the DeGLI paper ($\DIPDM_{cg}$).
\begin{table}
    \centering
        \scalebox{0.9}{
        \begin{tabularx}{\columnwidth}{lllllll}
        \toprule
        {} &    AP &    DM & DIP$_{cg}$ & DIP$_{cg}^{\mathrm{DM}}$ & DIP$_{vm}$ & DIP$_{vm}^{\mathrm{DM}}$ \\
        \midrule
        Iteration &  \textgreater{}100 &    43 &         12 &                       32 &          \textbf{5} &     \textbf{5} \\
        Time (s)  &  \textgreater{}3.14  &  1.36 &       1.43 &                     1.50 &        0.66 &     \textbf{0.54} \\
        \bottomrule
        \end{tabularx}
        }
    \caption{Average iteration number and runtime to reach $E_0 \leq 0.1$ on an NVIDIA V100 GPU, for on-dataset reconstruction (\cref{ssec:on-dataset}). $\DIPvm$ shows the best performance both as an initialization step before DM and a phase retrieval method in its own right.}
    \label{tab:runtime_to_e0}
\end{table}
\subsection{Off-dataset (Stanford Dogs)}
\vspace{-7px}
\label{ssec:off-dataset}
In the second scenario, we let the same reconstruction techniques compete on off-dataset simulated ptychographs, generated from images in the Stanford Dogs dataset \cite{datasetStanfordDogs}. We use the same ptychography settings as for training and the on-dataset scenario, except that we crop and resize the images to $64 \minitimes 64$ in contrast to MNIST's $28 \minitimes 28$. The bottom row of \cref{fig:errors_overview} shows the resulting error plots, and \cref{fig:images} shows reconstructed images. Both instances of DIP again show quick initial progress with respect to $E_0$, which is also clearly visible in the reconstructed images after the 10th iteration (\cref{fig:images}, second column). While the PSNR results are not as convincing, we would argue that both DIP instances seem to generalize surprisingly well to these structurally very different images. In particular, \cref{fig:images} shows that they are both able to eliminate ghost artifacts quickly, and successfully avoid the periodic artifacts that AP exhibits even after 100 iterations. With respect to $E_0$, $\DIPvm$ again comes out on top in this scenario.

\section{Conclusion}
\vspace{-2mm}
\label{sec:conclusion}
In this work we present a novel approach to the phase retrieval problem in ptychography. Based on a similar approach introduced for speech processing, we propose to augment conventional iterative algorithms with a complex convolutional neural network, such that the algorithm's output is refined by the network at each iteration. Evaluation on real-valued objects from image data shows an improvement in terms of convergence rate compared to conventional algorithms. Future extensions of DIP for complex-valued objects are planned. By evaluating on an unrelated dataset, we also show that the neural network does indeed learn the refinement task and does not merely adapt to the structure of training data. Our results indicate that the proposed method is also well-suited as an initialization step for other algorithms, thus allowing for an optimized trade-off between runtime and reconstruction quality.
\clearpage
\AtNextBibliography{\small}
\section{REFERENCES}
\label{sec:refs}
\atColsBreak{\vskip5pt}
\printbibliography[heading=none]
\end{document}